\documentclass[aps,pre,reprint,10pt,twocolumn,showpacs,nofootinbib,superscriptaddress,longbibliography]{revtex4-1}
\usepackage{epsfig}
\usepackage{graphics,graphicx,color,subfigure}
\usepackage{amssymb}
\usepackage{amsmath}
\usepackage{amsfonts}
\usepackage[utf8]{inputenc}
\usepackage{comment}
\usepackage{subfigure}
\usepackage[english]{babel}
\usepackage{color}
\usepackage[hidelinks,unicode=true]{hyperref}
\hypersetup{colorlinks=true,
linkcolor=blue,
urlcolor=blue,
 citecolor=blue,
pdfhighlight=/N
}%
\usepackage{microtype}
\usepackage[normalem]{ulem}

\newcommand{\lrangle}[1]{\langle{#1}\rangle}

\newcommand{\lbc}{\lambda_c}
\newcommand{\kmax}{k_{\mathrm{max}}}

\begin{document}
\title{Griffiths effects of the susceptible-infected-susceptible epidemic model on random power-law networks}
\author{Wesley Cota}
\email{wesley.cota@ufv.br}
\affiliation{Departamento de F\'isica, Universidade Federal de Vi\c cosa,
36570-000, Vi\c cosa, MG, Brazil}
\author{Silvio C. Ferreira}
\email{silviojr@ufv.br}
\affiliation{Departamento de F\'isica, Universidade Federal de Vi\c cosa,
36570-000, Vi\c cosa, MG, Brazil}
\author{G\'eza \'Odor}
\email{odor@mfa.kfki.hu}
\affiliation{MTA-MFA-EK Research Institute for Technical Physics and Materials Science, H-1121 Budapest, P.O. Box 49, Hungary}

\date{\today}

\begin{abstract}
We provide numerical evidence for slow dynamics of the
susceptible-infected-susceptible model evolving on finite-size random
networks with power-law degree distributions. Extensive simulations
were done by averaging the activity density over many realizations of
networks. We investigated the effects of outliers in both highly
fluctuating (natural cutoff) and non-fluctuating (hard cutoff) most
connected vertices. Logarithmic and power-law decays in time were
found for natural and hard cutoffs, respectively. This happens in
extended regions of the control parameter space
$\lambda_1<\lambda<\lambda_2$, suggesting Griffiths effects, induced
by the topological inhomogeneities. Optimal fluctuation theory
considering sample-to-sample fluctuations of the pseudo thresholds is
presented to explain the observed slow dynamics. A quasistationary
analysis shows that response functions remain bounded at $\lambda_2$.
We argue these to be signals of a smeared transition. However, in the
thermodynamic limit the Griffiths effects loose their relevancy and
have a conventional critical point at $\lambda_c=0$. Since many real
networks are composed by heterogeneous and weakly connected modules,
the slow dynamics found in our analysis of independent and finite
networks can play an important role for the deeper understanding of
such systems.
\end{abstract}

\maketitle

Quenched randomness in interacting dynamical systems causes non-trivial critical
behavior in nonequilibrium finite-dimensional
systems~\cite{vojta06,Vojta2006Rev,Munoz2010,Moretti2013,Noest1986,durrett}.
Spatial randomness can be introduced, among other ways, in the form of
dilution~\cite{adr-dic96,DeOliveira2008,vojta09} or nonuniformity of the control
parameters~\cite{vojta14c,durrett,Hooyberghs2004} or by topological
heterogeneity of the connectivity structure of the
interactions~\cite{Munoz2010,oliveira2,Juhasz2012,Odor2015,Moretti2013}. One of
the most noticeable effects of quenched (or quasistatic) disorder is the onset
of {dynamical
criticality, manifested in diverging correlation times and slow decays of the
order parameter in extended regions of the parameter space}, rid of fine
tuning~\cite{Vojta2006Rev}. This allows a potential for explaining widespread
observation of criticality, even without the assumption of self-organized
mechanisms~\cite{SOC}.

Extended criticality induced by quenched disorder is grounded on the
existence of rare regions (RRs), which are large, randomly occurring
patches that can linger for long times in a phase that differs from the
global state  of the system. Lets consider interacting dynamical
systems with active and inactive (absorbing) phases and a control
parameter $\lambda$ such that for $\lambda>\lambda_c$ the system is
globally active (supercritical) and for $\lambda<\lambda_0$ it is
inactive without long lived active RRs~\cite{Vojta2006Rev}. For
$\lambda_0<\lambda<\lambda_c$, the activity in RRs lasts for very long
(exponential in patch size) periods  but fluctuations unavoidably end
up the local activity due the finite size of the patches. Convolution
of low-probability RRs and exponentially long lifetimes results in a
slow dynamics with nonuniversal  exponents in the interval
$\lambda_0<\lambda<\lambda_c$ called Griffiths phase
(GP)~\cite{Griffiths1969,Noest1986}.

Complex networks constitute a fundamental theoretical framework to
describe substrates where many dynamical processes, as epidemics,
information, and transportation take
place~\cite{barrat2008dynamical,Dorogovtsev2008}. Heterogeneity
(disorder) is an intrinsic hallmark of complex networks manifested
through several forms of centralities~\cite{newman2010networks}. The
degree-centrality ranks among the most basic properties and is
statistically represented by the degree probability distribution $P(k)$
that a randomly selected vertex of the network has $k$
connections~\cite{Dorogovtsev2008}. Many networks observed in nature
have highly heterogeneous patterns of connectivity usually described
by a power-law (PL) degree distribution. These can be scale-free (SF)
networks~\cite{Albert2002}, characterized by heavy-tailed distribution
$P(k)$ with the ratio $\lrangle{k^2}/\lrangle{k}\gg \lrangle{k}$. Other
important measure of the dynamics on networks is the eigenvector
centrality~\cite{newman2010networks} associated with the principal 
eigenvector of the adjacency matrix defined as $A_{ij}=1$ if vertices
$i$ and $j$ are connected and 0 otherwise.

This intrinsic disordered nature of networks calls for analogues of GP
and RR phenomena. This issue has recently been investigated
\cite{Munoz2010,Juhasz2012} and GPs have been found in the contact
process (CP) \cite{ContactProcess} on finite-dimensional
networks. It was conjectured that GPs are not present in models on
infinite-dimensional, small-world graphs, where the average distance
between vertices increases logarithmically or slower with the network
size~\cite{Albert2002} as, for example, the case of random {PL}
networks. On the other hand, at models defined on hierarchical modular
structures, where the inter-module connectivity is weak, GPs were
reported~\cite{Moretti2013,Odor2015}.

The susceptible-infected-susceptible (SIS) epidemic
model~\cite{Pastor-Satorras2001}, in which infected vertices spontaneously heal
with rate 1 (fixing the time scale) and infect each of the susceptible nearest
neighbors with rate $\lambda$, is a paradigmatic example of {a} non trivial
dynamical process on complex networks. Differing from other dynamical processes
with transitions from active to inactive states, the SIS threshold is governed
by the activation of hubs, also called star subgraphs, and their mutual
reinfection through connected
paths~\cite{chatterjee2009,Mountford2013,boguna2013nature,ferreira2015collective}. As a consequence, the threshold is proved to be null in the infinite size limit for random networks with PL degree distribution $P(k)\sim k^{-\gamma}$, irrespective of the degree exponent $\gamma$~\cite{chatterjee2009}. \footnote{Reference~\cite{chatterjee2009} calls the model known in the physics literature as SIS of ``contact process''} Some interesting physical mechanisms behind the rigorous proof of Chatterjee and Durret~\cite{chatterjee2009} were used by Bogu\~n\'a \textit{et al}.~\cite{boguna2013nature} to unveil the nature of the epidemic threshold of the SIS model. Stars are graphs with $k\gg 1$ leaves connected to a center that can themselves sustain long-term epidemic activity if $k\gg 1/\lambda^2$. In PL networks the average highest degree diverges with the network size~\cite{mariancutofss}, implying that more and more hubs become active stars at finite values of $\lambda$. Due to the small-world property {of random PL networks}, the lifetime of hubs is large enough to permit infecting each other, sustaining epidemic activity in the network~\cite{boguna2013nature,ferreira2015collective}.

In finite random networks, the SIS dynamics is puzzling due to  the
highly fluctuating size and the number of stars realized in a network
sample~\cite{Ferreira12,Goltsev12,Lee2013,mata2014multiple}. Indeed,
the effective finite-size epidemic threshold is
finite~\cite{Ferreira12,boguna2013nature} since $k\ll 1/\lambda^2$ for
sufficiently small values of $\lambda$ and stars alone
cannot sustain a long-term activity. Several network realizations have
just a few vertices with degree much larger than the rest of network,
{hereafter} called outliers. Outliers can sustain localized epidemics, with
different activation thresholds for very long times, producing
multiple transitions~\cite{mata2014multiple}. Neglecting interactions
among hubs, Lee \emph{et al}.~\cite{Lee2013} predicted that the threshold
for an endemic phase, in which a finite fraction of infected vertices
is present, should take place at a finite value for $\gamma>3$, where
$\lrangle{k^2}$ is finite. According to Lee \emph{et al}.~\cite{Lee2013}, the
subcritical region is ruled by a GP, with an ultraslow (logarithmic)
decay of activity, in odds with the rigorous results of null threshold
for infinite PL networks~\cite{chatterjee2009}. This proposition was
supported by numerical simulations on very large, but finite, random
PL networks~\cite{mata2014multiple}. Finally, in finite PL networks, a vanishing epidemic threshold is predicted by the quenched
mean-field (QMF) theory~\cite{Wang03,Castellano10}, in which the full
connectivity structure of the network is included through the
adjacency matrix~\cite{Goltsev12}. In such models, Griffiths effects
were also shown in the localized phase for $\gamma>3$~\cite{odor14}.
RR effects, localization, and heavy-tailed dynamics have also been
shown in spreading models defined on weighted PL networks by
suppressing hub infection via disassortative weight schemes
\cite{Odor2012,odor13a}, in random networks \cite{odor13b,buono13}, or
in aging Barabasi-Albert graphs \cite{odor13b}.

Although localization effects were obtained in simulations on finite
networks~\cite{mata2014multiple}, the investigated systems were very
large ($\sim 10^8$ vertices), suggesting that these can be observable
and relevant in many unavoidably finite real networks. Aiming at a
deeper understanding of the intricate behavior of epidemic spreading
on finite-size  networks, we investigate the dynamics of the SIS model on
a large ensemble of PL networks, using extensive numerical
simulations. We show that the {averaging} over many independent graph
realizations exhibits a slow dynamics, analogous to GPs, in an
interval of control parameter $\lambda_1<\lambda<\lambda_2$. This region is delimited
by two transitions: The former is related to the activation of the
most connected hub of the network, while the latter is {related} to a
smeared phase transition~\cite{Vojta2006Rev}. Our results indicate that
this region shrinks {as the size of the network
increases and disappears} in the thermodynamic limit, implying the absence
of GPs. This is in agreement with the conjecture that finite
dimensionality is required for the existence of GPs~\cite{Munoz2010}.
Moreover, many real networks, as, for instance, brain connectomes, have
a modular organization, where modules are finite, heterogeneous, and
weakly connected~\cite{gallos12}. Slow dynamics has been reported in
models defined on hierarchical modular
networks~\cite{Moretti2013,Villages,Odor2015}. Therefore, performing
analysis over independent, finite networks can be useful to understand
these systems.

We have organized the paper as follows. The epidemic model and the
simulation methods are described in Sec.~\ref{sec:model}. Results
for the density decay and quasistationary simulations are presented
and discussed in Secs.~\ref{sec:time} and \ref{sec:qs},
respectively. An optimal fluctuation theory to explain the observed
slow dynamics is developed in Sec.~\ref{sec:opt}. Our concluding
remarks are presented in Sec.~\ref{sec:conclu}.

\section{Models and methods}

\label{sec:model}

The SIS model is defined as follows. Individuals lie in the vertices
of  a quenched network of size $N$ and can be in two states: infected
or susceptible. An infected individual $i$ becomes spontaneously
susceptible with rate  $1$, while a susceptible one turns to the
infected state with rate $\lambda n_i$, where $n_i$ is the number of
infected nearest neighbors of $i$. The dynamics is simulated on
networks obtained by the configuration model~\cite{newman2010networks}
with PL distribution $P(k)\sim k^{-\gamma}$, minimum  degree $k_0$ and
upper cutoff $\kmax$. Different cutoffs were investigated: Free
($\kmax=N$, strictly and is also called natural), hard
($\kmax=k_0N^{0.9/(\gamma-1)}$), and structural ($\kmax=\sqrt{N}$) cutoffs. The first one leads to degree distributions in
which  both the average and the standard deviation of the highly
fluctuating natural cutoff diverge as
$N^{1/(\gamma-1)}$~\cite{mata2014multiple}. Conversely, the second one
is engineered to render distributions without very large gaps in their
tails of the degree distribution{, since 
the factor $0.9$ guarantees that
$\sqrt{\lrangle{\kmax^2}-\lrangle{\kmax}^2}/\lrangle{\kmax} \rightarrow 0$ as $N\rightarrow\infty$~\cite{mata2014multiple}}.
The structural cutoff {is not fluctuating and} guarantees the absence of
degree correlations for $\gamma<3$ and becomes equivalent to the absence of a
cutoff for $\gamma > 3$ and $N\rightarrow\infty$~\cite{mariancutofss}. Graph
edges are generated randomly, forbidding multiple and self-connections. All
simulations were performed for $k_0=3$. Three ranges of the degree exponents
were considered separately: $\gamma>3$, for which localization is conjectured 
by the QMF theory~\cite{Goltsev12}; $\gamma<2.5$, where non-localized epidemic
spreading is predicted~\cite{Lee2013,Castellano12}; and $2.5<\gamma<3$ strict SF
regime, where localization is conjectured by the QMF theory~\cite{Goltsev12,Castellano12}.

The simulations were run using the modified Gillespie algorithm described in
Refs.~\cite{Ferreira12,mata2014multiple}. The number of infected vertices $n$
and the number of edges $S$ emanating  from them are computed and constantly
updated. With probability $1/(n+\lambda S)$ an infected vertex is randomly
chosen and cured. With the complementary probability, $\lambda S/(n+\lambda S)$,
an infected vertex $j$ is chosen with a probability proportional to its degree.
A vertex in the neighborhood of $j$ is chosen with equal chance and, if it is
susceptible, becomes infected, otherwise the simulation runs to next step.  The
time is incremented by $\Delta t = 1/(n+\lambda S)$ and the procedure is
repeated iteratively.
\begin{figure}[tbh!]
\centering
\includegraphics[width=0.9\linewidth]{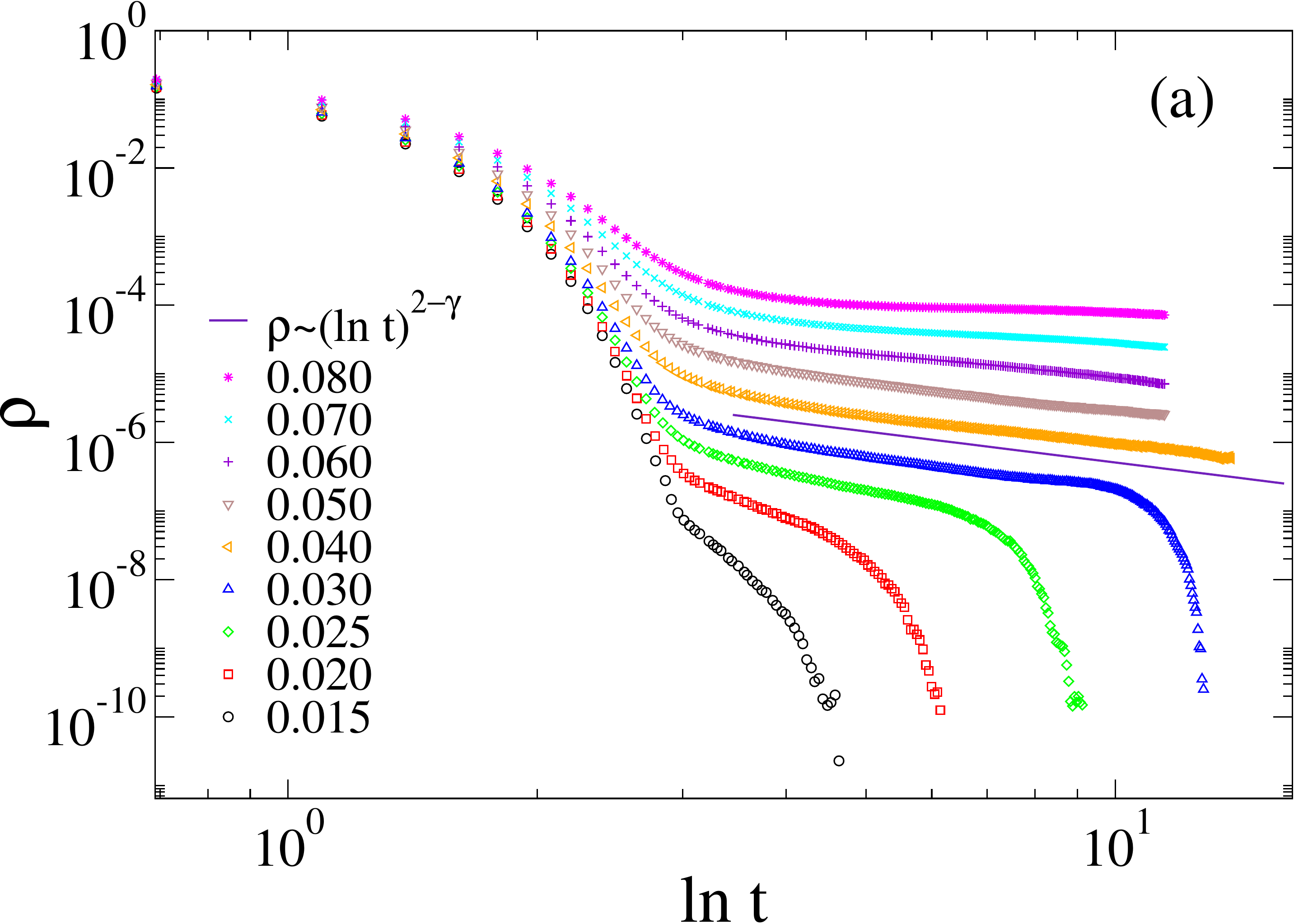}

\includegraphics[width=0.9\linewidth]{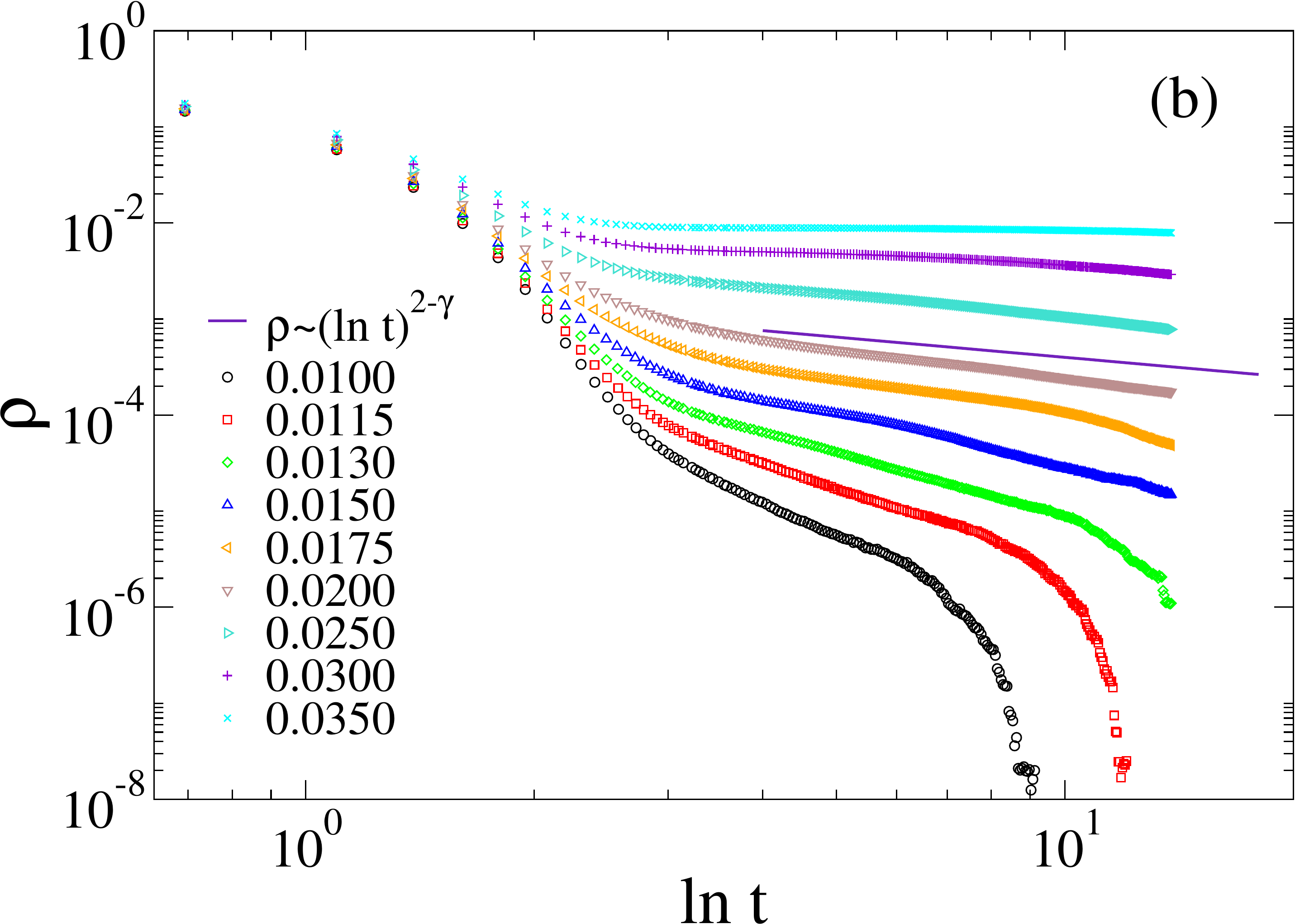}

\caption{Decay of density of infected vertices for networks with free upper
cutoff $(\kmax=N)$ using (a) $\gamma=3.5$ and (b) $\gamma=2.7$ for networks with sizes
$N = 10^7$ and $N=10^5$, respectively. The number of
samples were up to $\mathcal{N}=500$ and 4000 for $\gamma=3.5$ and 2.7,
respectively. Lines are predictions of the optimal fluctuation theory in
Sec.~\ref{sec:opt}. The values of $\lambda$ are indicated in the legends.}
\label{fig:g350free}
\end{figure}

We performed both standard and quasistationary (QS) analysis. In the former, the
decay from a fully infected initial state was investigated. In the latter, only
samples that did not visit the absorbing state are used to compute statistics
and the analysis was done in the (quasi) stationary regime~\cite{Marrobook}. We
used the improved QS method of Ref.~\cite{DeOliveira2005}, with parameters
similar to those given in Ref.~\cite{mata2014multiple}. Differently from
previous analysis of SIS on SF networks~\cite{mata2014multiple, Castellano10,
boguna2013nature, Ferreira12}, we aim at the average behavior of a large number $\mathcal{N}$ of
independent realizations. In the decay analysis, 10 to 100 independent  runs
were performed for each network, being the largest number of runs used for
smallest values of the infection rate.

\section{Time-dependent analysis}
\label{sec:time}

\subsection{{Free cutoff}}
\label{sec:free}

Simulations for networks with a free cutoff are shown in
Fig.~\ref{fig:g350free}. We observe an extremely slow logarithmic
decay in an extended  region of the control parameter $\lambda$.
For $\gamma=3.5$, the decay is very well fitted by 
\begin{equation}
 \rho\sim (\ln t)^{2-\gamma}.
 \label{eq:rholee}
\end{equation}
For $\gamma=2.7$, in the SF regime, we also see a logarithmic decay
$\rho\sim (\ln t)^{-\alpha}$, with a varying exponent $\alpha$ that is not quantitatively well described by Eq.~\eqref{eq:rholee}. The origin of the logarithmic
decay given by Eq.~\eqref{eq:rholee} is  related {to} the presence of the outliers in
the network and will be analytically explained {in
Sec.~\ref{sec:opt} using an optimal fluctuation theory.}

\subsection{{Hard cutoff}}
\label{sec:hard}

\begin{figure}[th!]
\centering
\includegraphics[width=0.805\linewidth]{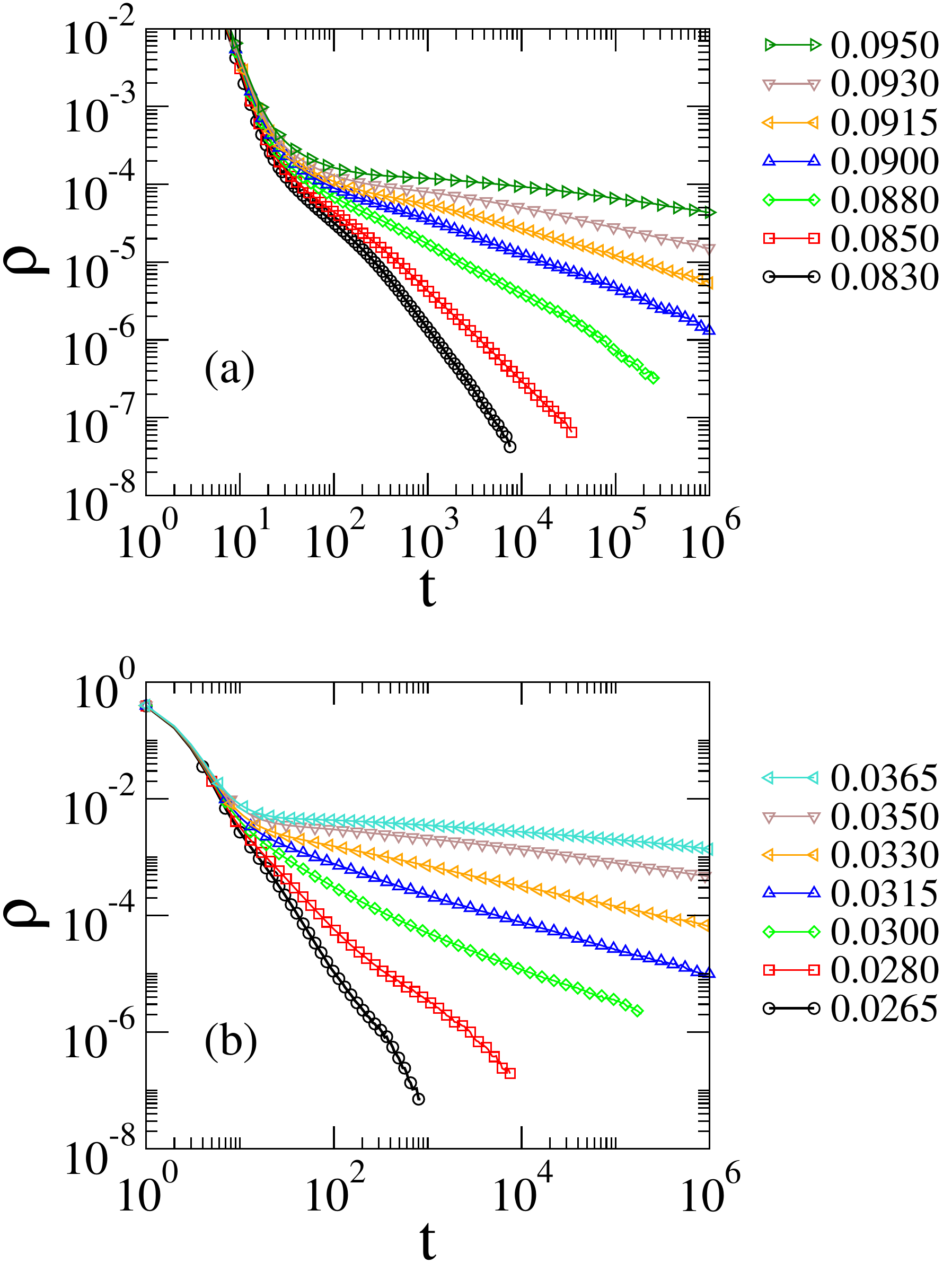}
\caption{Decay of density of infected vertices for (a) $\gamma=3.5$
and (b) $\gamma=2.7$  for networks of sizes $N=10^7$ and $10^5$,
respectively, using hard cutoff $k_c=k_0N^{0.9/(\gamma-1)}$. The
numbers of independent networks are $\mathcal{N}=500$ and 4000 for
$\gamma=3.5$ and $2.7$, respectively. The values of $\lambda$ are indicated in the legends.}
\label{fig:Rhog350g270hard}
\end{figure}

The localization of the QMF theory for $\gamma>2.5$ in concentrated
around the largest hub~\cite{pastor2016distinct}. So, the role
played by the hubs can be evidenced damping their number and
fluctuations. Evolution of the density of infected vertices for hard
cutoff of $P(k)$ is shown in Fig.~\ref{fig:Rhog350g270hard} for
$\gamma=3.5$ and 2.7. The data indicate PL decay with nonuniversal
exponents at long times for both degree exponents. Regressions fits:
$\rho\sim t^{-\alpha(\lambda)}$ at $\gamma=3.5$ resulted in $\alpha$
varying from 0.70 to 0.17 by increasing $\lambda$ from 0.088 to 0.095.
Similar range of exponents were found for $\gamma=2.7$, varying
$\lambda$ from 0.030 to 0.0365.

\subsection{{Structural cutoff}}
\label{sec:struc}
\begin{figure}[th!]
\centering
\includegraphics[width=0.805\linewidth]{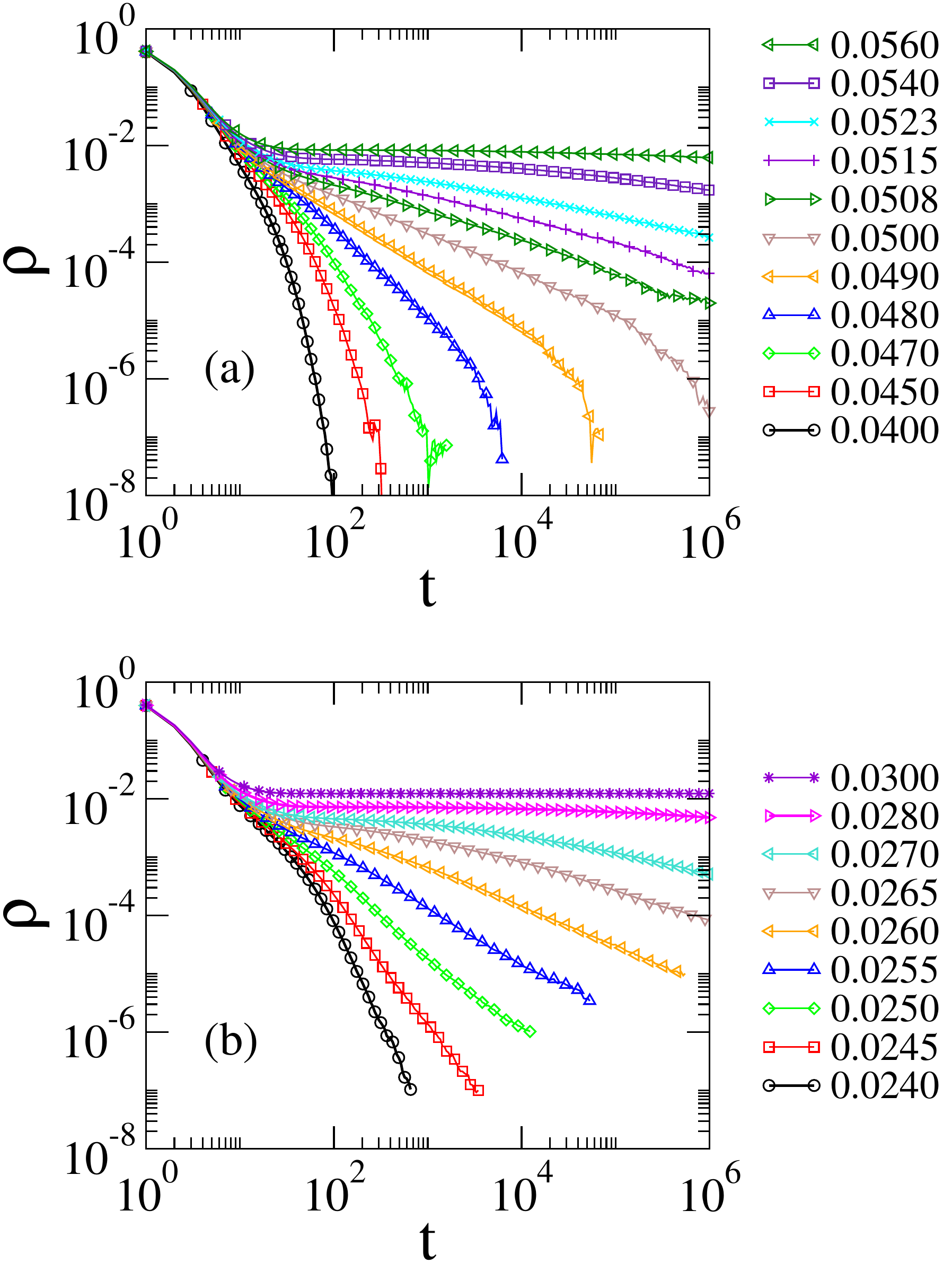}
\caption{Decay of density of infected vertices for (a) $\gamma=2.7$ 
and (b) $\gamma=2.3$  for networks of size $N=10^5$ using structural
cutoff $k_c=N^{1/2}$. The values of $\lambda$ are indicated in the legends.}
\label{fig:Rhog270g230ucm}
\end{figure}

We also simulated the density decay in networks with the structural
cutoff $k_c=N^{1/2}$ {with $\gamma<3$ since otherwise it is 
equivalent to the natural one}. This cutoff leads to the uncorrelated
configuration model (UCM)~\cite{Catanzaro2005}, that has been used in many
analyses of SIS on SF networks~\cite{Ferreira12, boguna2013nature,
Castellano10,Castellano12, mata2013pair,Lee2013}. Power-law decays in time are
still observed, but the extended region is reduced compared with
hard cutoffs, see Fig.~\ref{fig:Rhog270g230ucm}.

\subsection{{Sample-to-sample fluctuations}}
\label{sec:samp}

\begin{figure}[th!]
	\centering
	\includegraphics[height=0.99\linewidth]{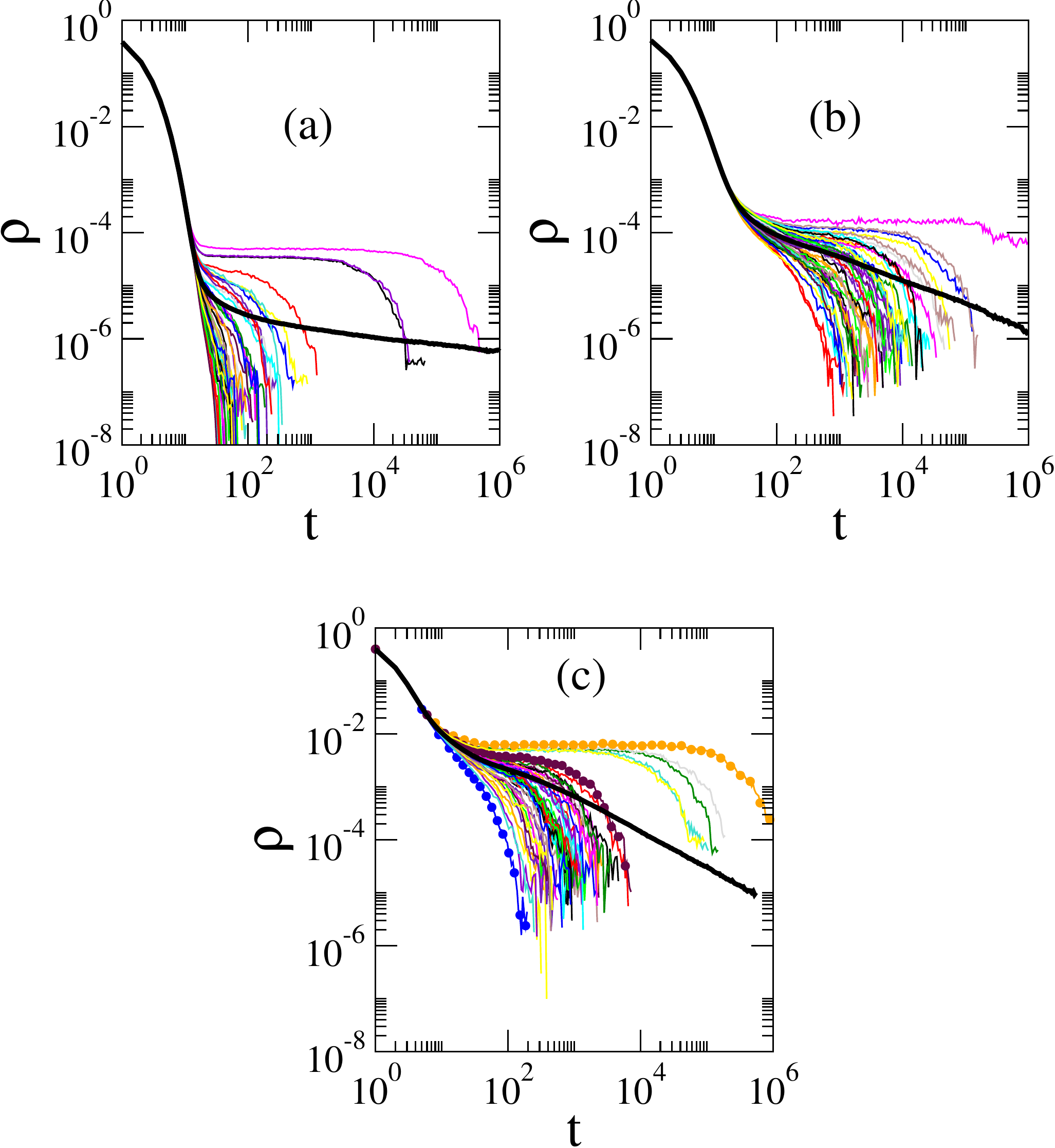}
	\caption{Sample-to-sample fluctuation of the evolution of SIS on PL networks: (a)
		$\gamma=3.5$, $N=10^7$, free cutoff, and $\lambda=0.04$; (b) $\gamma=3.5$, hard
		cutoff $k_c=k_0N^{0.9/(\gamma-1)}$, $N=10^7$, and $\lambda=0.09$; (c)
		$\gamma=2.3$, structural cutoff $k_c=N^{1/2}$, $N=10^5$, and $\lambda=0.026$.
		Curves for 50 independent networks are shown. Thick lines represent 
		the average of $\mathcal{N}=500$ and 2000 samples for $\gamma=3.5$ and $2.3$, respectively.} \label{fig:50samples}
\end{figure}

The origin of the slow decay is the sample-to-sample fluctuations, rather than
occurrence of rare regions in the same network. Figure~\ref{fig:50samples} shows
the decay of the density for 50 networks with all parameters fixed to values for
which slow decays are observed in the averaged curves. One can see that several
curves are subcritical, while others behave super-critically, evolving to a
metastable stationary density value before falling in the absorbing state.
{Numerically, we  observe pseudo thresholds and, consequently, all quantities of
interest with wide distributions whose relative variance does not decrease with
the network size within the region of Griffiths effects. This means that the
quenched disorder is relevant within the dynamical critical region analogously
to a lack of self-averaging~\cite{Aharony1996}, where having many finite samples is not
equivalent to averaging over a single large network~\cite{Dorogovtsev2008}; note that the dynamical critical region where disorder is relevant
diminishes as network size increases and disappears in the thermodynamical
limit, as discussed later.}

There exist two main mechanisms for this large variation, the leading one
depends on $\gamma$ and the cutoff used. For a free cutoff, the size of the
largest hubs fluctuates greatly. So, the presence of outliers, creating local
active domains, determines if the dynamics levels off to a quasi steady state in
the simulation of that sample. In {the} cases of hard or structural cutoffs,
the tails of the degree distributions fluctuate little {and}
the leading mechanism is the variation of the overall heterogeneity of the
network, which can be measured by the average degree of the nearest neighbors of
the vertices $k_{nn}$~\cite{Alexei}. For the structural cutoff case this becomes
$k_{nn}=\lrangle{k^2}/\lrangle{k}$~\cite{mariancutofss}, whose inverse provides
a very precise estimate of the SIS epidemic threshold for
$\gamma<2.5$~\cite{Ferreira12,mata2013pair}. Density decay for networks with
$\gamma=2.3$, using a structural cutoff, are shown in
Fig.~\ref{fig:50samples}(c). Three samples highlighted with symbols possess
$\lrangle{k}/\lrangle{k^2} = $ 0.0270, 0.0243, and 0.0235, and the larger values the
lower densities. These values must be compared with the infection rate
$\lambda=0.026$ used in all samples. We see that samples for which
$\lrangle{k}/\lrangle{k^2}\lesssim \lambda$ are supercritical and those where
$\lrangle{k}/\lrangle{k^2}\gtrsim \lambda$ are subcritical. An optimal
fluctuation theory to explain this slow dynamics is presented in
Sec.~\ref{sec:opt}.

\subsection{{Finite-size analysis}}
\label{sec:finite}

\begin{figure}[tbh]
\centering
\includegraphics[width=0.99\linewidth]{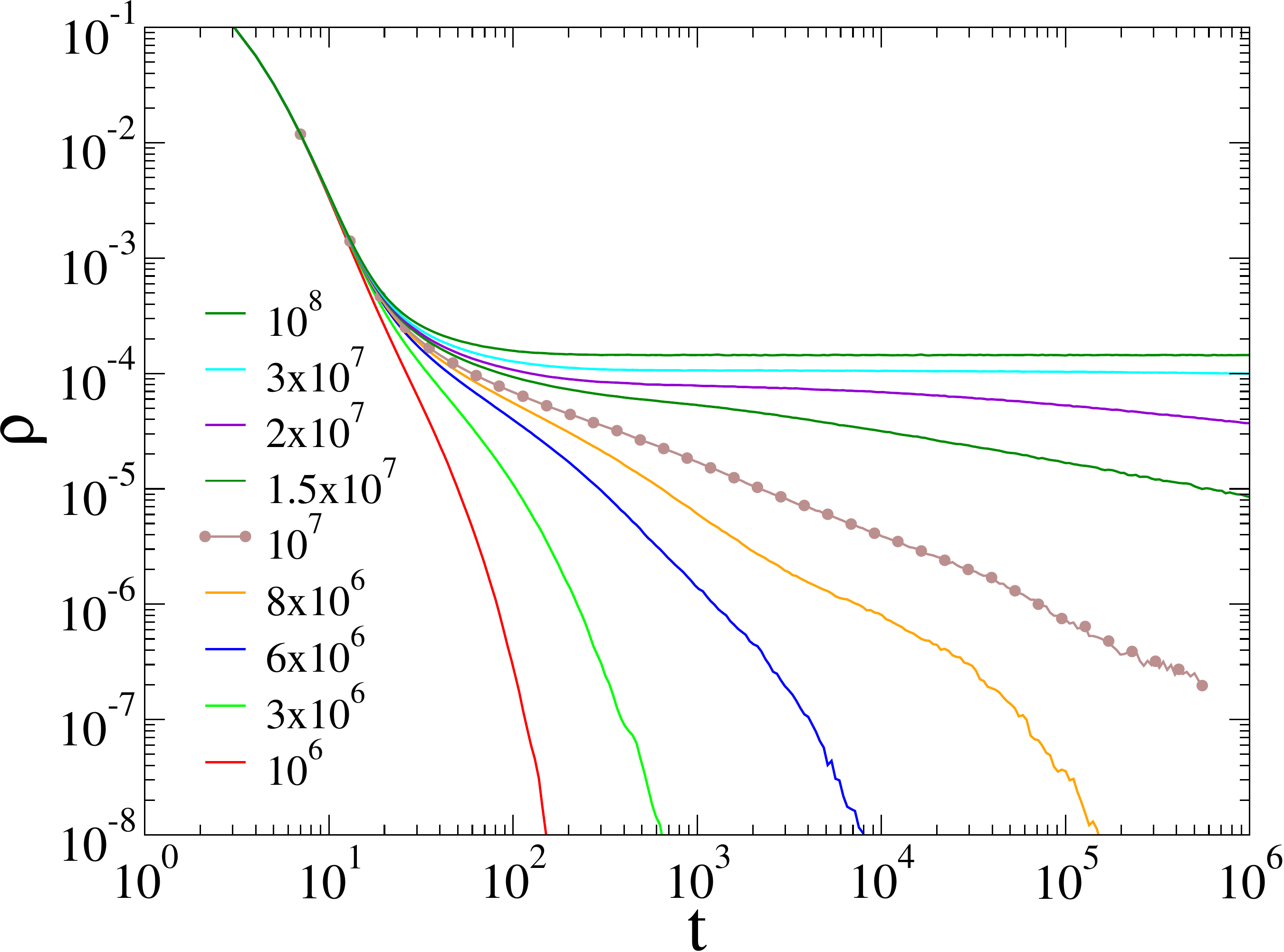}
\caption{Finite-size analysis of the density decay against time for
networks with $\gamma=3.5$ and hard cutoff $k_c=k_0N^{0.9/(\gamma-1)}$
for infection rate $\lambda=0.088$. The number of network samples is
100 for the two largest sizes and 500 for the others. The network sizes are indicated in the legend.}
\label{fig:lb0p088hardg3p5}
\end{figure}

The slow dynamics observed in the ensemble averages is not a genuine
Griffiths singularity, since it is not triggered by the slowly decaying
RRs; thus we can expect that these effects disappear in the
thermodynamic limit.  This conjecture is confirmed in
Fig.~\ref{fig:lb0p088hardg3p5}, where we show the density of infected
vertices against time for different sizes for a fixed infection rate {$\lambda=0.088$}.
We see that the dynamics is deeply subcritical, an exponential decay
of activity, for $N=10^6$. For size $N=10^7$ a PL regime can be
observed but, finally, a saturation to a constant plateau develops at
$N=10^8$. The disappearance of the PL regimes is mainly associated
with the shift of the epidemic threshold towards zero as the size
increases~\cite{Ferreira12,mata2013pair}. The threshold drops from
approximately 0.12 for $N=10^6$ to 0.075 for $N=10^8$. Similar
finite-size effects were observed in the CP on weighted
trees~\cite{Odor2012} and the same mechanism shown here is probably
also present {there}. The CP also exhibits strong
finite-size dependence of the thresholds, approaching the asymptotic
value only at exceeding large networks~\cite{Ferreira2011cp,Mata2014}.
Moreover, the range of $\lambda$ where PLs are observed decreases as
$N\to\infty$ (see also Sec.~\ref{sec:qs}), thus Griffiths effects {disappear} in
the thermodynamic limit. 

\section{Optimal fluctuation theory}
\label{sec:opt}

The hypothesis {drawn} in Sec.~\ref{sec:time}, in which the slow decay
is originated from sample-to-sample fluctuations of the effective thresholds 
in finite-size networks can be put in a mathematical ground 
by approximating the sample average with an integral
\begin{equation}
\bar{\rho} = \int_0^\lambda d\lbc \rho(\lambda,\lbc)\mathcal{P}(\lbc)e^{-t/\tau(\lambda,\lambda_c)}.
\label{eq:rhoav}
\end{equation}
Here $\rho(\lambda,\lbc)$ is the quasi-static density as {a} function of
$\lambda>\lambda_c$ ($\rho\equiv 0$ for $\lambda<\lambda_c$),
$\tau(\lambda,\lbc)$ is the lifetime of the dynamical
processes and $\mathcal{P}(\lbc)$ is the probability density 
that a randomly selected sample has a threshold at $\lambda_c$.

{We assume that in a free {cutoff} network with 
$\gamma>2.5$ the activation happens at the most 
connected hub~\cite{Castellano12}.
Consider a star subgraph, centered on the vertex of maximal degree
$\kmax$, which forms an independently activated domain in a network 
with $N$ nodes. Using QMF theory, the threshold in such {a} star} 
graph is $\lambda_c \sim 1/\sqrt{\kmax}$~\cite{Ferreira12,mata2013pair}. 
The density in a star of size $k$ is\footnote{The actual density 
in a star must increase a $\rho\sim \lambda^{\beta_\mathrm{star}}$, 
where $\beta_\mathrm{star}>\beta_\mathrm{QMF}=1$.} $\rho_\mathrm{star}\approx\lambda$ for 
$\lambda \gtrsim \lambda_c$, implying that 
\begin{equation}
\rho=\frac{\lambda \kmax}{N}.
\label{eq:rhostar}
\end{equation}
The lifespan of the activity in a star in case of SIS dynamics is~\cite{boguna2013nature} 
\begin{equation}
\tau \simeq \tau_0 \exp(a \lambda^2 \kmax),
\label{eq:taustar}
\end{equation}
where $a$ and $\tau_0$ are constants. 
Finally, the probability of a given threshold is
$\mathcal{P}(\lbc) d\lbc = \Pi(\kmax)d\kmax$, where 
\begin{equation}
\Pi(\kmax)\simeq N\exp(-c N \kmax^{-\gamma+1}) \kmax^{-\gamma}.
\label{eq:Pi}
\end{equation}
is the probability of the 
largest degree to be $\kmax$ in a PL network with $N$ vertices~\cite{mariancutofss}. 
Here $c$ is a constant depending on $P(k)$.
Plugging Eqs.~\eqref{eq:rhostar}-\eqref{eq:Pi} into Eq.~\eqref{eq:rhoav},
we obtain
\begin{equation}
\bar{\rho}\sim \lambda \int_{1/\lambda^2}^\infty \kmax^{-\gamma+1} \exp(-cN\kmax^{-\gamma+1}) \exp(-t/\tau)d\kmax.
 \label{eq:rhoav2}
\end{equation}
If $1/\lambda^2\gg N^{1/(\gamma-1)}$, then the first exponential
suppresses the integral and a standard subcritical phase with 
exponential decay is expected. For $1/\lambda^2\lesssim N^{1/(\gamma-1)}$, the first
exponential is approximately 1. After an integration by parts this
integral can be easily evaluated using the saddle-point method to return
$\bar{\rho}\sim (\ln t)^{2-\gamma}$, exactly the result of
Eq.~\eqref{eq:rholee}. This decay is the same found by
Lee \textit{et al}.~\cite{Lee2013}, in a theory of SIS dynamics for
infinite PL networks, with non-interacting hubs.
This predicts ultraslow decay instead of a stationary
endemic state, contradicting the exact result of the null epidemic
threshold for SIS irrespective of $\gamma$~\cite{chatterjee2009}. Why
do our simulations match this theory? The assumption that stars 
form independent domains of activity is incorrect in principle, 
since the lifetime of epidemics on stars can be sufficiently large 
to permit mutual infection of hubs, even if they are not directly 
connected due to the small-world property~\cite{boguna2013nature,ferreira2015collective}. 
However, several stars that contributed to the average epidemic activity 
in our simulations are observed in different realizations of
networks and are thus actually independent.

In the SF regime at $\gamma=2.7$ the QMF still predicts localization
(see Appendix), but there is a high probability that several activated
hubs occur in the same network sample even if their size is finite.
Thus neglecting the multiplicity of activated hubs as well as the
interaction among them~\cite{boguna2013nature} is not a quantitatively
accurate approximation but it is able to capture the essentially
logarithmically slow dynamics observed in simulations.

In case of hard cutoff, we do not know the form of
$\mathcal{P}(\lambda_c)$. Since the fluctuations of $\lambda_c$ depend
on global properties of the networks, we assume their distribution to
be Gaussian, with width $\sigma(N)$ and centered at $\lambda_0(N)$,
tending to a delta function at $\lambda=0$ as  $N\rightarrow\infty$,
in conformity with numerics. Less is know about the lifespan.
Numerically, we have data consistent with $\tau\sim
\exp[a(\lambda-\lambda_c)^2]$, where $a(N)$ is some function
increasing with the size that we could not determine precisely.
Plugging these forms into Eq.~\eqref{eq:rhoav} and using the saddle-point approximation to solve the integral we found $\bar{\rho}\sim
t^{-1/2a\sigma^2}$. This is a nonuniversal power law, in agreement
with the density decay simulations, as far as $a\sigma^2$ is a
nonuniversal constant.

\section{Quasi-stationary analysis}
\label{sec:qs}

\begin{figure}[thb]
\centering
\includegraphics[height=0.49\linewidth,angle=90]{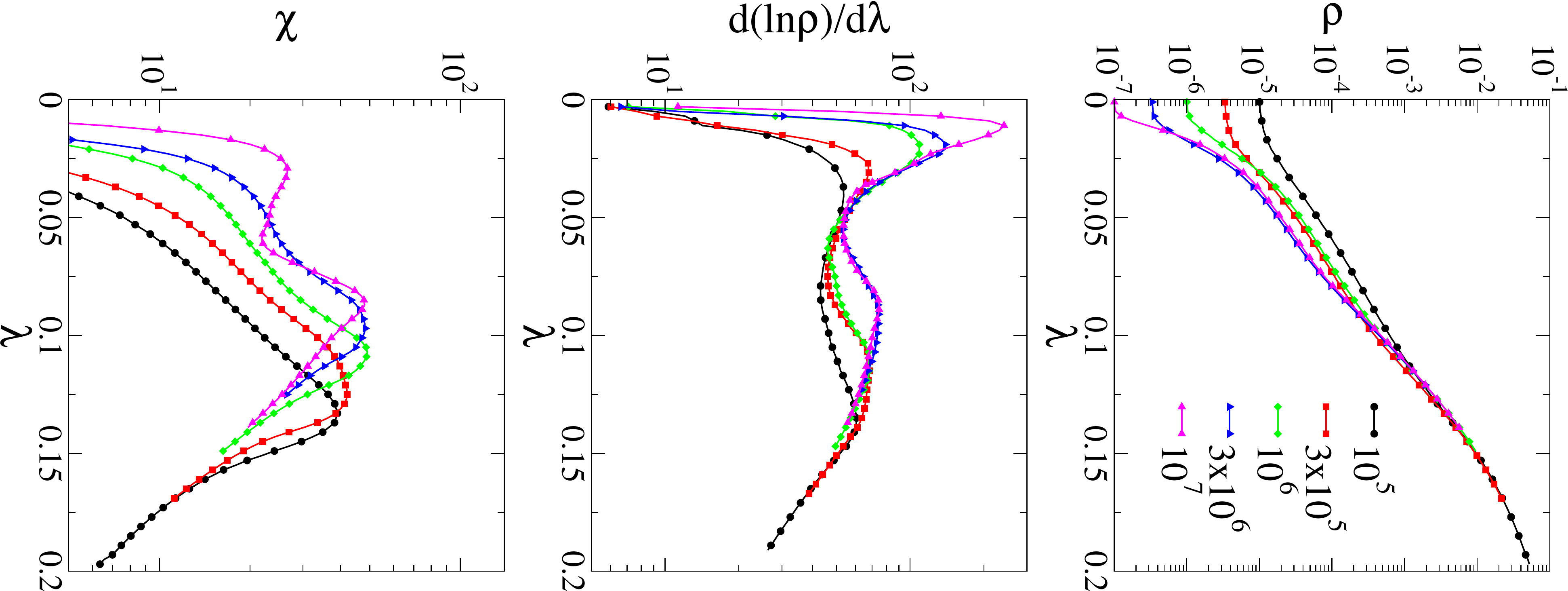}
\includegraphics[height=0.49\linewidth,angle=90]{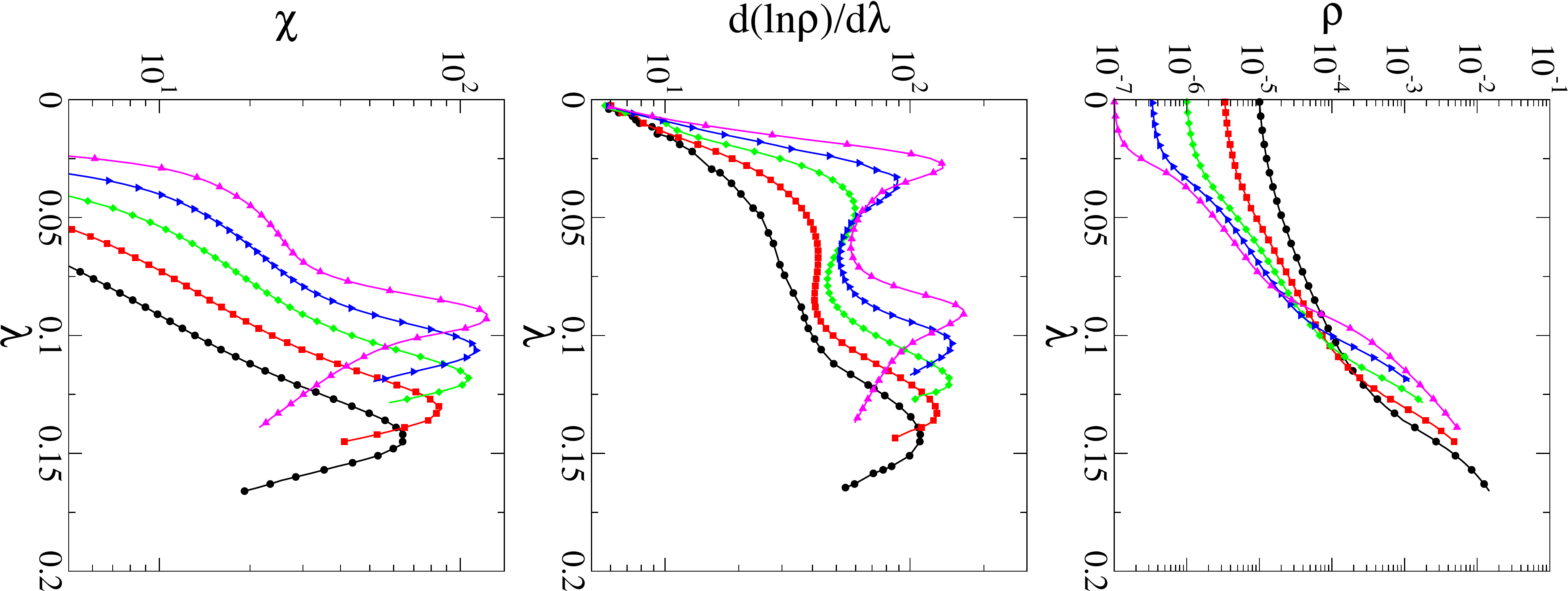}
\caption{Finite size analysis of the QS state for networks with $\gamma=3.5$.
The [(a) and (b)] QS density $\rho$,  [(c) and (d)] log-derivative of $\rho$, and [(e) and (f)]
dynamical susceptibility are shown for free and hard cutoffs, respectively. The number of samples were at least
$\mathcal{N}=50$. The network sizes are indicated in the legend.}
\label{fig:Rho_vs_lambda_free}
\end{figure}

The stationary state of the SIS model on a single finite network with
degree exponent $\gamma>3$ was characterized by multiple transitions 
as $\lambda$ is varied due to the independent activation 
of different regions with different thresholds~\cite{mata2014multiple}. 
One transition was associated with the activation of the most 
connected vertex with a vanishing threshold of the QMF theory 
$\lambda_1\sim 1/\sqrt{\kmax}$.
A second transition was associated with the mutual activation of
hubs~\cite{boguna2013nature} and took place at a threshold
$\lambda_2$. A third transition,
          that occurs at another finite threshold, was also found. 
This represents the collective activation of the network~\cite{ferreira2015collective}. 
Again, we tackle this problem by averages over a large ensemble.

Figures ~\ref{fig:Rho_vs_lambda_free}(a) and (b) compares the QS
density against $\lambda$ for degree exponent $\gamma=3.5$ with
either hard or free cutoffs for networks of different sizes. The
average QS density is a double sigmoid, a non-monotonically increasing
function of $\lambda$, which indicates two phase transitions. At
a standard clean critical point, the logarithmic
derivative of the QS density scales as~\cite{Dickman2006}
\begin{equation}
\left. \frac{d\ln \rho}{d\lambda}\right|_{\lambda_c}\sim L^{1/\nu_\perp},
\end{equation}
where $L$ {is the} system size and $\nu_\perp$ is a critical exponent
associated with the divergence of the correlation length. This
log-derivative can also be used to identify multiple transitions in
epidemic spreading in networks~\cite{mata2014multiple} in association
with the dynamical susceptibility
$\chi=N(\lrangle{\rho^2}-\lrangle{\rho}^2)/\lrangle{\rho}$
\cite{Ferreira12}. The latter quantifies the relative fluctuations of
the order parameter as shown in Fig.~\ref{fig:Rho_vs_lambda_free}.

The log-derivative of the density is shown in Figs. ~\ref{fig:Rho_vs_lambda_free}(c) and (d), while the
susceptibility is shown in Figs. ~\ref{fig:Rho_vs_lambda_free}(e) and (f).
Averages over the ensemble of networks wipe out the multiple
transitions of single networks, leading to two observable transitions
at thresholds $\lambda_1$ and  $\lambda_2>\lambda_1$. These correspond
to the peaks of the log-derivatives and
susceptibility; the latter is less evident for $\lambda_1$. Notice,
that the double transition identified with the hard cutoff in the
log-derivative analysis starts to emerge as a shoulder in the
susceptibility curves of the largest size investigated. The threshold
at $\lambda_1$ can clearly be seen in the susceptibility curves for
natural cutoff only for very large sizes and is manifested as a
shoulder for the other cases, including hard cutoffs.

For the free cutoff, a QS density $\rho_{\mathrm{free}}\gg 1/N$, the
minimal value allowed in a QS simulation, is observed in the interval
$\lambda_1<\lambda<\lambda_2$. This resembles a smeared phase
transition~\cite{Vojta2006Rev} and the interval coincides with the 
region, where Griffiths effects are found in the density decay analysis.  
We attribute this smearing to the presence or absence of outliers in different
samples. In the case of hard cutoff, the suppression of outliers
leads to a weaker smearing with a density $1/N\ll
\rho_{\mathrm{hard}} \ll \rho_{\mathrm{free}}$. In a standard
smeared phase transition, patches having high-enough dimension can
exhibit ordering transition independently. In principle outliers, 
represented by stars, are high-dimensional objects, which could be 
activated independently. So the basic ideas of smeared transitions 
could be fulfilled. However, outliers plus their neighbors provide a vanishing fraction 
of the network and give a vanishing contribution to the 
global density in the thermodynamic limit. Thus they generate
a finite-size effect.

\begin{figure}[tbh]
\centering
\includegraphics[width=0.99\linewidth]{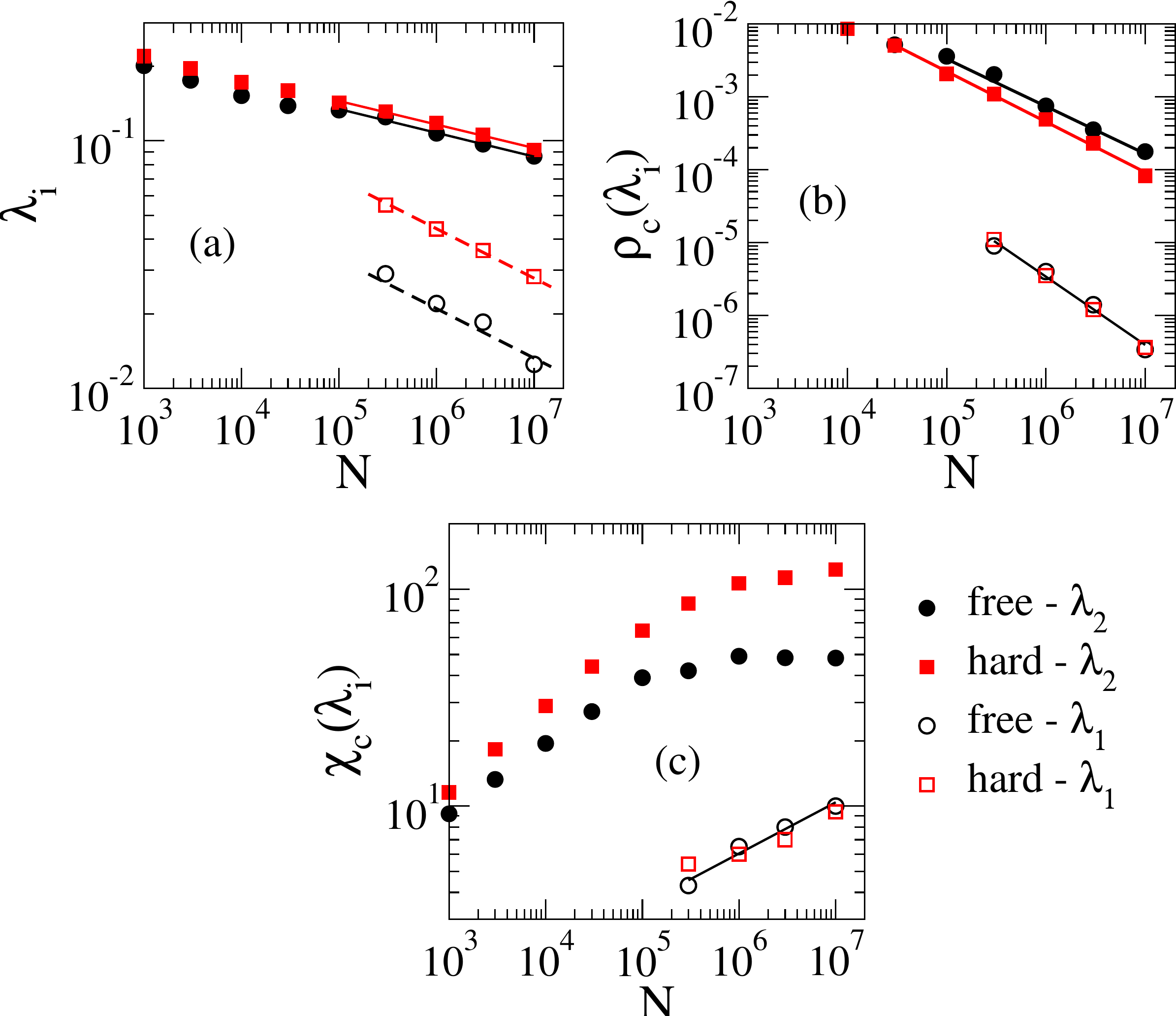}
\caption{Critical quantities against networks size for $\gamma=3.5$:
(a) threshold, (b) critical density, and (c) critical susceptibility at
$\lambda=\lambda_i$. Solid lines are PL regressions while the dashed
lines are $\lambda_1\sim N^{-0.2}$ and $N^{-0.18}$
predicted by the QMF theory for free and hard cutoffs, respectively.}
\label{fig:crit}
\end{figure}

The finite-size analysis at thresholds $\lambda_{1,2}$ are shown in
Fig.~\ref{fig:crit}. The left threshold, determined via the
log-derivative, decays consistently with QMF theory and, for
$\gamma>2.5$, scales as $\lambda_1\sim 1/\sqrt{\kmax}\sim
N^{-0.5/(\gamma-1)}$ or $N^{-0.45/(\gamma-1)}$ for natural and hard
cutoffs, respectively. This coincides with the activation of the star
graph centered at the most connected vertex of the
network~\cite{Castellano10}. The position of the right peak of the
susceptibility, which agrees with the right one of log-derivative
curves, goes slowly to zero as $\lambda_2(N)\sim N ^{-0.09}$ in the
investigated interval while the density evaluated at $\lambda_2$
follows a power law $\rho_c(\lambda_2)\sim N^{-0.65}$. These scaling
laws are, within uncertainties, independent of the cutoff.
However, both response functions, log-derivative and susceptibility,
saturate with the size, confirming a smeared transition at
$\lambda_2$. The susceptibility at $\lambda_1$, for free cutoff,
increases as $\chi_c\sim N^{0.23}$, which is again consistent with the
activation of the most connected vertex: For a star with $\kmax$
leaves we have approximately $\chi\sim (\kmax)^{0.55}$ (using data
from Ref.~\cite{mata2013pair}) and given that $\kmax\sim N^{1/(\gamma-1)}$,
we obtain an exponent, which is close to 0.23, observed in the
simulations.

The peaks at finite values of $\lambda$, observed in single
networks~\cite{mata2014multiple}, are wiped out and are not evident
when averages are done.\footnote{ In Ref.~\cite{mata2014multiple}
networks with up to $N=10^8$  vertices were simulated while here  we
analyzed until $N=10^7$ and performed larger ensemble averaging. So
our results do not definitely discard other transitions for higher
sizes but no indications of them were observed in the investigated
size range.}

Analyzing the behavior for $N=10^7$ {with} free cutoff and $\gamma = 3.5$, we found $\rho
\sim \lambda^\beta$, where $\beta\approx 2.8$ with
$\lambda_1<\lambda<\lambda_2$. Running the dynamics only on the star
centered on the most connected vertex by permanently immunizing the
rest of the network, we found $\beta_\mathrm{star}\approx 2.0<\beta$. This means
that the mutual activation of hubs is relevant in the interval
$\lambda_1<\lambda<\lambda_2$, leading to an exponent larger than that
of a single star centered on the most connected vertex. The estimate
$\beta\approx2.8$ for $\gamma=3.5$ is inside the rigorous bounds found by
Chatterjee and Durret~\cite{chatterjee2009}:
$\gamma-1<\beta<2\gamma-3$.

\section{Conclusions}

\label{sec:conclu}
Random, scale-free networks exhibit strong, quenched inhomogeneities,
and therefore rare region effects can be expected to play an important
role. To see rare regions of arbitrary sizes we should simulate
arbitrarily large system sizes or by the standard way of
approximations we do sample averages over many independent network
realizations. The latter way is not equivalent to the former one in
scale-free networks. In models defined on networks with infinite
topological dimensions{,} a recent hypothesis states that Griffiths
phases cannot exist~\cite{Munoz2010} {and another} important result for
infinite-dimensional networks with power-law degree distributions is
that SIS does not exhibit a phase transition at finite
$\lambda$~\cite{chatterjee2009}. Real networks, on the other hand, can
be very large, but are always finite. Therefore, a numerical analysis on
different sizes is of great importance.

Here we present extensive simulations on networks generated with the
configuration model using free (fluctuating) and structural or {hard}
(nonfluctuating) degree cutoffs. We focused on statistics over a large ensemble
of networks. Contrary to the results obtained on single network realization,
where multiple transitions were reported~\cite{mata2014multiple}, we observe
that the network ensemble averaging exhibits Griffiths effects in an extended
region of the control parameter $\lambda_1(N)<\lambda<\lambda_2(N)${, which
	diminishes as network size increases and disappears in the thermodynamical limit.
	These Griffiths effects are due to sample-to-sample fluctuations, producing non-self-averaging within the shrinking critical dynamical region, rather than the
	existence of RRs of actual Griffiths phases.}
We also observe the occurrence of a smeared transition, with saturated
fluctuations of the order parameter at $\lambda_2$. Our findings  can be
relevant if we consider independent realizations as graphs occurring in a
sequence of uncorrelated, time-dependent networks at a given time and we measure
quantities in the long-time average. Alternatively{,} such results can describe the
behavior of systems in which power-law degree distribution in  modules make up
a very weakly coupled network.

More specifically for free cutoff networks, we found an asymptotic
logarithmic decay of density  in time in the interval
$\lambda_1(N)<\lambda<\lambda_2(N)$. Here $\lambda_1 \sim
1/\sqrt{\kmax}$ is associated with the activation of the most connected
vertex~\cite{Castellano10} and $\lambda_2$ describes the mutual
activation of hubs that leads to an endemic phase of the
network~\cite{boguna2013nature,ferreira2015collective}. Both
thresholds go to zero at the infinite size limit with
$\lambda_2/\lambda_1\rightarrow\infty$, implying that for finite sizes
there exists a detectable extended interval with Griffiths
effects. The logarithm decay is explained by an optimal fluctuation
theory for networks where fluctuating pseudo-critical points are
considered. For structural{ and hard cutoffs} the Griffiths effects are
weaker, resulting in nonuniversal power-law density decay tails within
a smaller range of the control parameter. We attribute this to the
lack of outliers, which causes weaker sample-to-sample heterogeneity
and less fluctuating pseudo-critical points.

The finite-size analysis shows that the transition peaks move
to zero by increasing the size and the Griffiths effects are replaced by
a conventional critical point behavior characterized by $\beta\approx 2.8$ for degree exponent $\gamma=3.5$.

\begin{acknowledgments}
This work was supported by the funding agencies CNPq, CAPES, and
FAPEMIG (Brazil) and the Hungarian research fund OTKA (Grant No.
K109577). We thank Robert Juh\'asz  for fruitful comments and
discussions. G.\'O. thanks the Physics Department at UFV, where part
of this work was done, for its hospitality. S.C.F. thanks discussions
with Romualdo Pastor-Satorras during visits to UFV supported by the program 
\textit{Ci\^encia sem Fronteiras} - CAPES (Grant No.  88881.030375/2013-01).
\end{acknowledgments}

\vspace{0.2em}
\appendix

\section{Spectral analysis for \texorpdfstring{$\gamma=2.7$}{γ = 2.7}}
\label{sec:app}

\begin{figure}[th!]
\centering
\includegraphics[width=0.89 \linewidth]{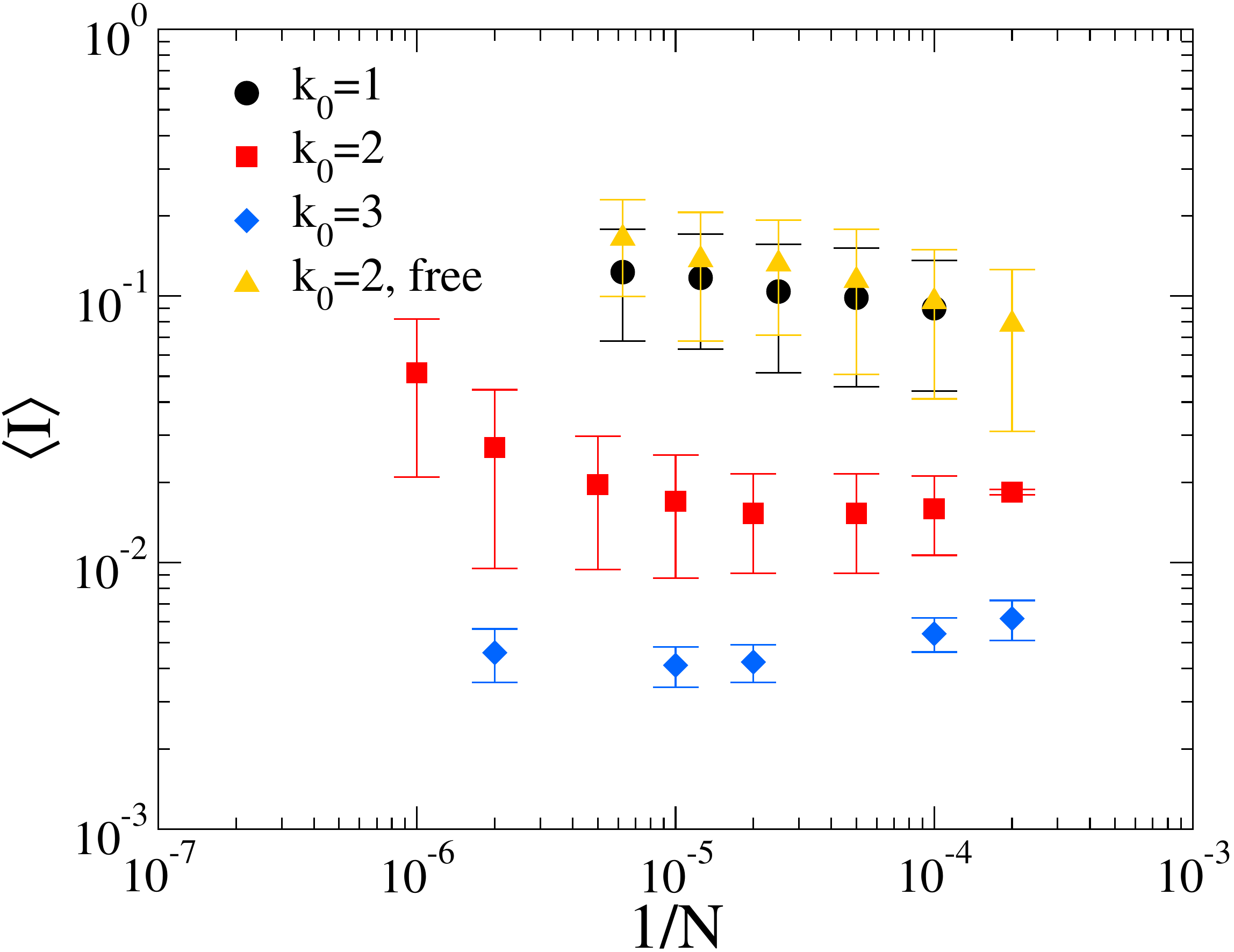}
\caption{Finite-size scaling of IPR for $\gamma=2.7$, 
$k_0=1,2,3$, and structural and $k_0=2$ with free cutoff.
\label{IPR28FSS}}
\end{figure}

We have also tested whether the QMF theory provides localization for $2.5 <
\gamma < 3$, since earlier numerical results suggested a localization
transition at $\gamma=3$ \cite{odor14}. We performed spectral analysis
of the SIS on UCM networks with $k_0=1,2,3$ structural and $k_0=2$
with free cutoff as described in Ref.~\cite{odor14} at $\gamma=2.7$ for
$N=5000 - 10^6$. We diagonalized the $A_{ij}$ matrix which describes
the evolution of activity probabilities $\rho_i(t)$ of node $i$ in the
QMF approach
\begin{equation}
\label{qmfsis}
\frac{d\rho_i}{dt} = -\rho_i + \lambda (1-\rho_i)\sum_{j=1}^N A_{ij} \rho_j~.
\end{equation}
The localization in the active steady state can be quantified by
calculating the inverse participation ratio (IPR) of the principal
eigenvector  $\mbox{\boldmath$e$}(y_1)$, related to the largest
eigenvalue of the adjacency matrix as~\cite{Goltsev12}
\begin{equation} \label{defIPR}
I(N) \equiv \sum_{i=1}^{N} e_{i}^{4}(y_1) \ .
\end{equation}
This quantity vanishes as $1/N$ in the case of homogeneous
eigenvector components, but remains finite as $N \to \infty$, if the
activity is concentrated on a finite fraction of nodes. The average
values $\langle I\rangle$  were determined for $100 - 1000$
independent networks for each parameter value. The finite size
analysis (Fig.~\ref{IPR28FSS})  for $k_0=1$ shows a clear monotonic
increasing tendency of $\langle I\rangle$ as $N\to\infty$. In case of a
$k_0 >1$, the mean IPR values decrease first and cross
over very slowly to an increase for $N>10^5$. For free cutoff the
graph generation is more difficult and slow, but again one can observe
a monotonic increase of $\langle I\rangle$ up to $N=160~000$.
Therefore, we see numerical evidence that epidemic activity of the 
QMF theory is localized at $\gamma = 2.7$ as expected, 
in general, for $\gamma>2.5$ \cite{Goltsev12}.

\bibliographystyle{abbrv}


%

\end{document}